\documentstyle[psfig]{texas}

\begin{document}

\title{Relativistic Jet Stability and Structure from the Alfv\'en
                 Point Outwards}

\author{Philip E. Hardee$^{1}$, Alexander Rosen$^{1}$, Philip A. Hughes$^{2}$ and G. Comer Duncan$^{3}$}

\address{(1) University of Alabama \\
	     Department of Physics \& Astronomy,
             Tuscaloosa, AL 35487 USA \\ }

\address{(2) University of Michigan \\
 	     Astronomy Department,
             Ann Arbor, MI 48109 USA \\ }

\address{(3) Bowling Green State University \\
             Department of Physics \& Astronomy,
             Bowling Green, OH 43403 USA \\ }

\begin{abstract}

The amplitude of jet distortions and accompanying pressure and velocity
fluctuations resulting from Kelvin-Helmholtz instability of three
dimensional relativistic jets are explored. The effect of instability
on jets as they accelerate from sub- to super-Alfv\'enic speeds is
explored and a relativistic stabilization mechanism for
trans-Alfv\'enic jets is proposed.  The level to which asymmetric
instabilities on supermagnetosonic relativistic jets will grow is
predicted theoretically and a Doppler boosted ``apparent'' emissivity
is computed.  Effects due to helically twisted filamentary structure
produced by asymmetric modes of instability should be readily
observable on relativistic jets.

\end{abstract}

\section{Introduction}

There is compelling evidence for relativistic highly collimated jets in
both galactic \cite{lb96} and extragalactic objects \cite{caw91}.  In
the galactic ``superluminals'' GRS 1915+105 \cite{mr94} and GRO
J1655-40 \cite{hr95} \cite{tetal95} the observed proper motions
indicate jet flow speeds on the order of 92\% of light speed.  Among
the extragalactic superluminals, observed motions indicate jet flow
speeds as large as 99.9\% of light speed, e.g., 3C~345 \cite{zcu95}.
It is quite likely that all jets produced by black holes in binary star
systems \cite{mr95} or in AGN's \cite{b94} \cite{b95} are initially
relativistic.

Recent jet acceleration and collimation schemes require dynamically
strong magnetic fields close to the central engine.  Numerical studies
\cite{mpl96} \cite{ops97} \cite{op97} \cite{rukcl97} show that the jets
created in this fashion pass through slow magnetosonic, Alfv\'enic, and
fast magnetosonic critical points.  The ultimate jet velocity may
depend on the configuration of the magnetic field \cite{megpl97}, and
the jets accelerate up to asymptotic speeds that may be only a few
times the Alfv\'en speed at the Alfv\'en point -- at the Alfv\'en point
the jet speed equals the Alfv\'en speed \cite{cam97}. This basic
acceleration and collimation process may be the same for all classes of
objects that emit jets \cite{l97}.

Jets are susceptible to the Kelvin-Helmholtz instability.   The K-H
instability of three-dimensional (3D) jets with purely poloidal or
purely toroidal magnetic fields \cite{r81} \cite{ftz81} \cite{fj84}
\cite{brfk89}, and of jets containing force-free helical magnetic
fields \cite{ac92} \cite{a96} has been extensively investigated in the
supermagnetosonic limit. This instability can lead to the development
of highly organized helical structures and ultimately to disruption of
highly collimated flow as mass is entrained into the jet from the
external environment \cite{rhcj99}.  In general, spatial or temporal
growth rates associated with the K-H instability are found to increase
as the magnetosonic Mach number decreases provided the jet is
super-Alfv\'enic.  Unlike purely fluid flows which are unstable when
subsonic, the poloidally magnetized jet is predicted to be nearly
completely stabilized to the K-H instability when the jet is
sub-Alfv\'enic \cite{hr99}.  Here in \S~2 and \S~3 we consider the
types of organized structure and stability properties that exist on
supermagnetosonic and trans-Alfv\'enic non-relativistic jets,
respectively.  In \S~4 and \S~5 we consider the stabilizing effects of
relativistic jet speeds in the trans-Alfv\'enic and supermagnetosonic
regimes.

\section{The Supermagnetosonic Non-Relativistic Jet}

Structures that arise on jets can be considered to consist of Fourier
components of the form 
$$
f_1(r,\phi,z)=f_1(r)exp[i(kz \pm n \phi - \omega t)]
$$
where flow is along the $z$-axis, and $r$ is in the radial direction
with the flow bounded by $r=R$. $k$ is the longitudinal wavenumber, $n$
is an integer azimuthal wavenumber, for $n > 0$ the wavefronts are at
an angle to the flow direction, the angle of the wavevector relative to
the flow direction is $\theta = tan(n/kR)$, and $\pm n$ refers to wave
propagation in the clockwise and counterclockwise directions around the
jet circumference.  $n =$ 0, 1, 2, 3, 4, etc.  correspond to pinch,
helical, elliptical, triangular, rectangular, etc. ``normal'' mode
distortions of the jet, respectively.  For a jet with uniform density,
temperature and axial magnetic field (top hat profile) and a uniform
external medium, the propagation and growth or damping of the Fourier
components is described by a dispersion relation \cite{bk91}.  When the
flow is supermagnetosonic, each normal mode, $n$, contains a growing
single ``surface'' wave and multiple ``body'' wave solutions that
satisfy the dispersion relation (cf., Figure 2).
\begin{figure} [h!]
\centering
\hspace*{-3.5cm}
\vspace*{-18cm}
\psfig{figure=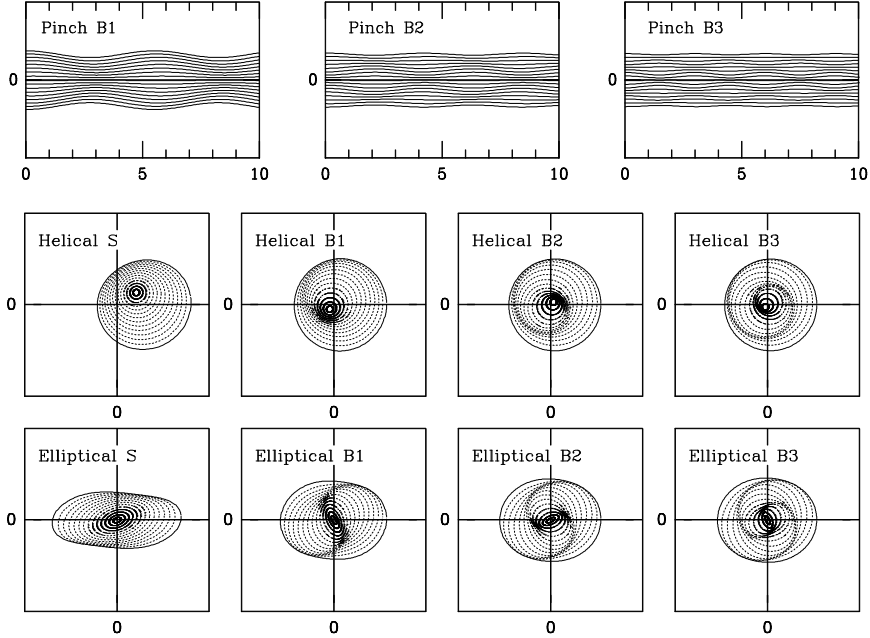,width=21cm,height=27cm}
\caption{Axial cross sections associated with maximum pinch body mode (B1, B2, B3) displacements and transverse cross sections associated with maximum helical and elliptical surface (S) and body mode (B1, B2, B3) displacements.}\label{fig1}
\end{figure}
The growth rate of normal mode surface and body waves has a maximum
that is inversely proportional to the magnetosonic Mach number at a
wavelength that is approximately proportional to the magnetosonic Mach
number.  The body wave modes exist only for wavelengths less than some
maximum wavelength.

Figure 1 shows the maximum displacements associated with asymmetric
surface wave modes at the fastest growing wavelength and of symmetric
and asymmetric body wave modes at the maximum unstable wavelength \cite{hcr97}
\cite{hrhd98}.  The body modes show a reversal in fluid displacement at
``null'' surfaces interior to the jet surface.  For example, the first
pinch body mode has a null displacement surface (fluid does not move
radially) at jet center.  The second pinch body mode shows an
additional null displacement surface between jet center and jet
surface.  The third pinch body mode shows two null displacement
surfaces between jet center and jet surface.  Pressure fronts inside
the jet become more oblique to the jet axis as the body wave mode
number increases.  The radial displacement associated with an
asymmetric ``surface'' wave mode $n > 0$ is approximately given by
$\xi_{r,n}(r)\approx \xi _{r,n}^s(r/R)^{n-1}$ where  $\xi _{r,n}^s$ is
the surface amplitude \cite{h83}.   Inside the jet surface the
displacement is rotated in azimuthal angle relative to displacement at
the jet surface in addition to changes in the displacement amplitude.
Fluid displacements associated with the asymmetric body waves show a
much larger azimuthal rotation and have larger amplitudes in the jet
interior relative to the jet surface than for the surface wave
\cite{hcr97}.  Only the helical surface and body waves lead to jet
displacement off the initial axis.  Jet distortions comparable to the
amplitudes shown in Figure 1 have been observed in non-relativistic 3D
MHD numerical simulations \cite{hcr97}.

\section{The Trans-Alfv\'enic Non-Relativistic Jet}

When a jet is sub-Alfv\'enic the helical and higher order ($n > 0$)
surface modes are stable, but rapidly destabilize when the jet becomes
super-Alfv\'enic.  The stability and behavior of the normal mode
solutions can be investigated analytically in the limit $\omega
\rightarrow 0$.  In this limit the real part of the pinch mode ($n=0$)
surface wave solution becomes \cite{hr99}
$$
\omega /k \approx u\pm \left\{ \frac 12\left( V_A^2+\frac{
V_A^2a_{j}^2}{a_{ms}{}^2}\right) \pm \frac 12\left[ \left( V_{A}^2+\frac{
V_{A}^2a_{j}^2}{a_{ms}{}^2}\right) ^2-4\frac{V_A^4a_{j}^2}{a_{ms}{}^2}
\right] ^{1/2}\right\} ^{1/2} ~. 
$$
In the equation above $V_A = (B^2/4 \pi \rho_{j})^{1/2}$, $a_{j}$ is
the jet sound speed, and $a_{ms} \equiv (a_{j}^2+V_A^2)^{1/2}$ is the
magnetosonic speed. The imaginary part of the solution is vanishingly
small in the low frequency limit. These solutions are related to fast
($+$) and slow ($-$) magnetosonic waves propagating with ($u+$) and
against ($u-$) the jet flow speed $u$, but strongly modified by the
jet-external medium interface. Numerical solution of the dispersion
relation reveals that a growing solution is associated with the
backwards moving (in the jet fluid reference frame) solution related to
the slow magnetosonic wave and the pinch mode surface wave is unstable
on sub-Alfv\'enic and super-Alfv\'enic jets.
 
When a jet is sub-Alfv\'enic the helical and higher order ($n>0$)
surface modes are stable \cite{brfk89} \cite{hr99} and have an outwards
moving purely real solution given by
$$
\omega /k\approx u+V_A ~. 
$$
On the super-Alfv\'enic jet in the limit $\omega \rightarrow 0$ all
higher order modes ($n>0$) have surface wave solutions given by
$$
\omega/k \approx \frac {\eta}{1+\eta}u  
\left\{ 1\pm i
\frac{\left[ 1 - (1+\eta)V_A^2/u^2\right]^{1/2}}{\eta^{1/2}}
\right\} ~.
$$
where the jet to external medium density ratio $\eta \equiv \rho
_{j}/\rho _{x}$. Wave growth corresponds to the plus sign.  Note that
in the dense jet limit, i.e., $\eta \rightarrow \infty $, this
expression reduces to  $\omega /k\approx u\pm V_A$, and thus the
surface waves are related to Alfv\'en waves propagating with and
against the jet flow speed but strongly modified by the jet-external
medium interface. The unstable growing solution is associated with the
backwards moving (in the jet fluid reference frame) wave. The surface
wave speed in the observer frame, $(\omega /k)_{Real}$, and the growth
rate, $(\omega /k)_{Imag}$, are strongly dependent on the density
ratio.  Higher jet density is stabilizing.

Numerical simulations of low density 3D MHD jets \cite{hr99} verify
that ``light'' jets destabilize rapidly near to the Alfv\'en point.  In
the simulations, the observed rapid development of instability leads to
rapid mass entrainment and slowing of the ``light'' jets in the
trans-Alfv\'enic region before the jets can become supermagnetosonic.

\section{Trans-Alfv\'enic Relativistic Jet Stabilization}

The rapid development of instability at the Alfv\'en point and
subsequent mass entrainment and slowing of ``light'' jets can be
overcome.  To see what is required for enhanced stability of
the helical and higher order normal modes we consider the asymmetric
normal mode solutions to the fully relativistic MHD dispersion relation
in the limit $\omega \rightarrow 0$~\cite{h99a}:
$$
\omega/k \approx {\gamma^2\eta \over 1+\gamma^2[1+(V_A/\gamma c)^2]\eta} u  
\left\{1 \pm i
{[1 - (1+\eta)(V_A/ \gamma u)^2-(\eta/\gamma^2)(V_A/c)^4]^{1/2} \over \gamma \eta^{1/2}}\right\}
$$
where $\eta \equiv W_{j}/W_{x}$, $V_A \equiv (B^2/4\pi W_j)^{1/2}$,
$\gamma$ is the Lorentz factor, and the enthalpy $W \equiv \rho +
({\Gamma \over \Gamma - 1})(P/c^2)$ where $\Gamma$ is the adiabatic
index.  A positive value for the imaginary part of the frequency
indicates instability. Note that the Alfv\'en speed can be
superluminal.  When $[1 - (1+\eta)(V_A/ \gamma
u)^2-(\eta/\gamma^2)(V_A/c)^4] < 0$ the jet is stable to the asymmetric
normal modes, and in the limit $\eta \rightarrow \infty $ it can be
shown that
$$
\omega /k \approx {u\pm v_w \over 1 \pm (uv_w/c^2)} {\rm ~~and~~} v_w = {V_A \over (1 + V_A^2/c^2)^{1/2}} ~.
$$
Thus, helical and higher order surface waves are related to Alfv\'en
waves and, of course, the Alfv\'en waves propagate at less than
lightspeed.  The unstable solution is associated with the backwards
moving (in the jet fluid reference frame) wave.  When unstable the wave
speed in the observer frame, $(\omega /k)_{Real} \approx [\gamma^2\eta
/(1+\gamma^2 \eta )]u$, increases and the growth rate is reduced as
$\gamma \eta^{1/2}$ increases.  Thus, relativistic jets are effectively
heavier and can pass through the Alfv\'en point with less rapid
development of instability.  These effectively heavier jets should be
much less susceptible to slowing by mass entrainment \cite{rhcj99}.

\section{Supermagnetosonic Relativistic Jet Structure}

\subsection{Theory \& Axisymmetric Numerical Simulation Compared}

Jets that survive the trans-Alfv\'enic region without serious mass
entrainment and slowing will become supermagnetosonic with continued
expansion.  The amplitude to which normal mode structures can develop
on supermagnetosonic relativistic jets can be deduced from numerical
simulations of relativistic jets \cite{dh94} \cite{hrhd98}.  In the
simulations the jets are perturbed by strong pressure waves
driven into the jet at approximately constant spatial interval by large
scale vortices in the surrounding cocoon.  The cocoon vortices induce
oblique pressure disturbances in the jet at approximately the
relativistic Mach angle.  This oblique disturbance is at the
appropriate angle to couple to the third pinch body mode (B3 in Figure
1).

Figure 2 shows normal mode solutions to the dispersion relation
appropriate to a relativistic jet simulation \cite{dh94} (Lorentz
factor $\gamma = 5$, Mach number $\gamma M_j \approx 10.5$ where $M_j
\equiv u/a_j$, and the jet resides in a hot tenuous cocoon where $M_c
\equiv u/a_c \approx 1.5$)  for surface (S) and the first three body
waves (B1, B2, B3) of the pinch, helical and elliptical normal modes.
These solutions are typical of supermagnetosonic jet flows.  With the
exception of the surface pinch mode all wave modes have a maximum in
the growth rate (dashed lines).  The body wave modes have a longest
unstable wavelength (determined from the real part of the wavenumber
where the imaginary part goes to zero) which is slightly shorter than
the longest allowed wavelength (determined from the wavenumber in the
limit $\omega \rightarrow 0$).
\begin{figure} [h!]
\centering
\hspace*{0.0cm}
\vspace*{-4.5cm}
\psfig{figure=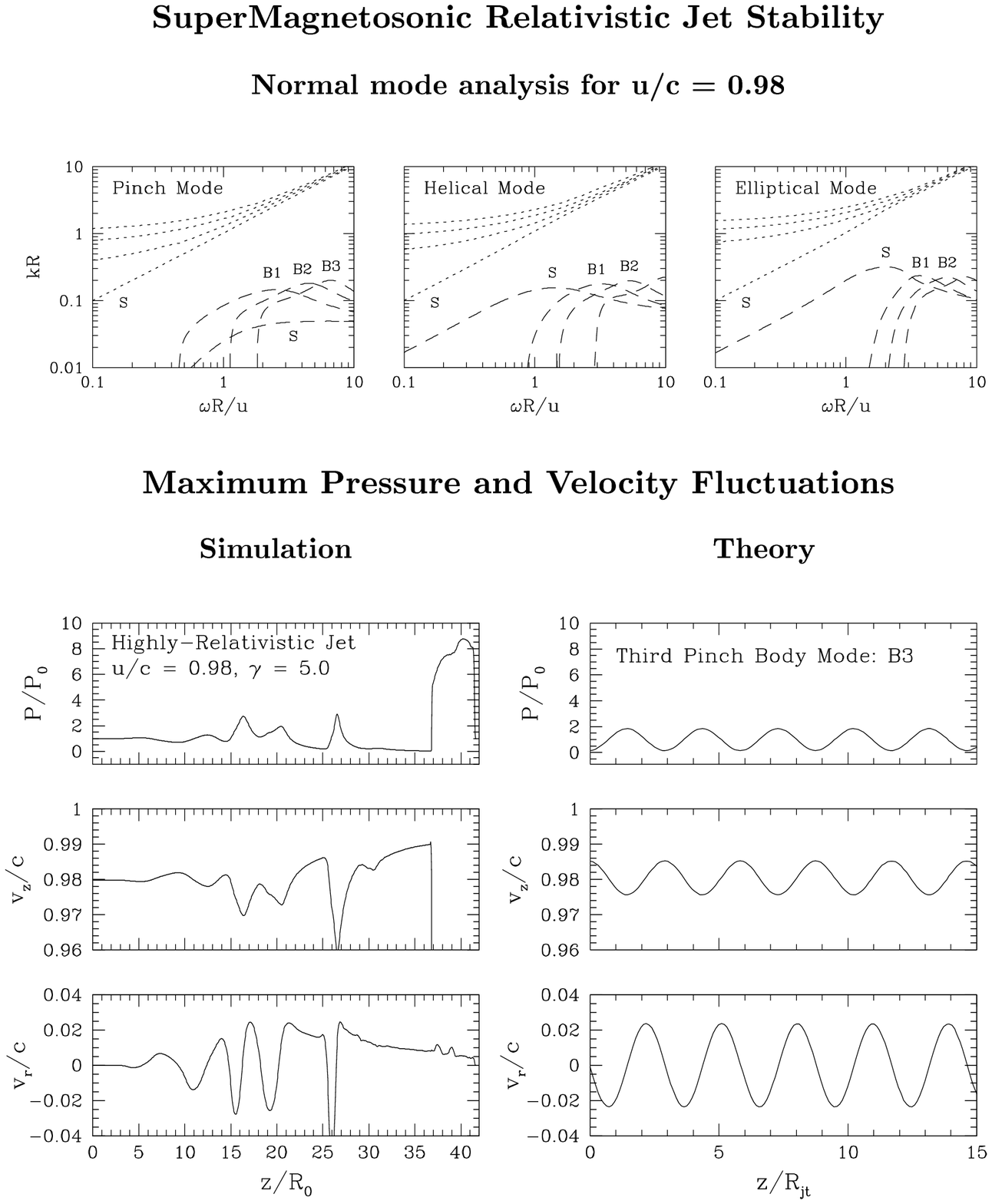,width=16cm,height=20cm}
\caption{Solutions to the dispersion relation for the highly
relativistic numerical simulation (top panels).  The dotted (dashed)
lines give the real (imaginary) part of the wavenumber as a function of
the angular frequency.  Pressure and velocity fluctuations seen in the
numerical simulation just off the jet axis, and pressure and velocity
fluctuations associated with the B3 pinch body mode just off the jet
axis are shown in the bottom panels.}\label{fig2}
\end{figure}
Figure 2 also contains a comparison between pressure, axial, and
transverse (radial) velocity fluctuations seen in the numerical
simulation with theoretically derived oscillations associated with the
B3 pinch body mode at the longest unstable wavelength.  The comparison
indicates that a maximum theoretical pressure fluctuation, $2 > P/P_0 >
0$, successfully reproduces the observed maximum transverse (radial)
velocity oscillations seen in the numerical simulation far behind the
jet front, i.e., $15 < z/R_0 < 20$.

\subsection{Predicted Asymmetric Relativistic Jet Structure}

The maximum pressure fluctuation criterion ($2 > P/P_0 > 0$) found from
axisymmetric numerical simulations can be used to predict the cross
section distortions and asymmetric structural features that will appear
on 3D supermagnetosonic relativistic jets.
\begin{figure} [h!]
\centering
\hspace*{-1.25cm}
\vspace*{-4.0cm}
\psfig{figure=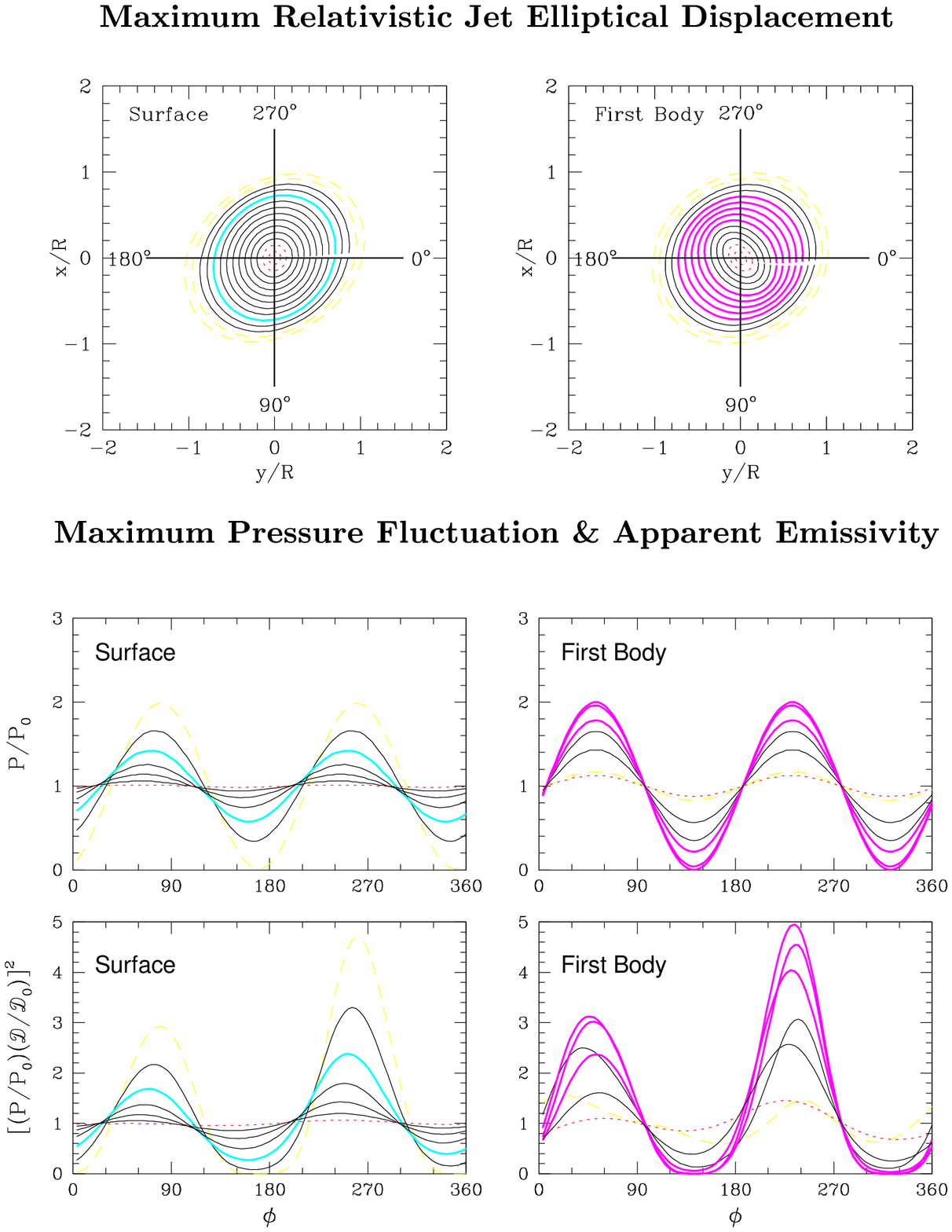,width=16cm,height=20cm}
\caption{Maximum transverse cross section elliptical surface and first
body wave distortions (top panels), and the accompanying pressures and
``apparent'' emissivities around the displacement contours (bottom
panels). The dashed (heavy solid, dotted) lines in the cross sections correspond to the dashed (heavy solid, dotted) line(s) in the pressure and apparent emissivity plots.}\label{fig3}
\end{figure}
In Figure 3 (top panels) we plot the maximum cross section distortions
for the elliptical surface wave mode and first elliptical body wave
mode at the maximum growth rate and longest unstable wavelength,
respectively, where we have used the dispersion relation solutions
appropriate to the relativistic jet shown in Figure 2, i.e., $u/c =
0.98$ and $\gamma = 5.0$.  Comparison between these distortions and
those allowed on non-relativistic jets (see Figure 1) shows a
considerable reduction in the allowed distortion.  In general, we have
found that the allowed distortion decreases as the relativistic Mach
number increases at wavelengths near to the maximum growth rate
(surface wave) or between the longest unstable wavelength and the
maximally growing wavelength (body waves).  At longer wavelengths the
surface wave elliptical distortion to the cross section can be larger
than that shown in Figure 3 \cite{h99b}.

High pressure regions accompanying the distortions along with the
Doppler boosting factor $$ D \equiv \left\{\gamma [1-(v/c) cos \theta
]\right\}^{-1} $$ and an angle $\theta = 1/\gamma + \delta \theta$
where $\delta \theta$ represents the perturbation to the initial flow
angle, $\theta = 1/\gamma$, with respect to the line of sight, allows
us to construct an ``apparent'' synchrotron emissivity $P^2D^2$.  For
the elliptical distortions shown in Figure 3, $| \delta \theta | < 0.1
{\rm ~radian}$, and we also have found that $| \delta \theta | \propto
1/\gamma $. The curves of pressure and ``apparent'' emissivity shown in
Figure 3 are taken around the displacement contours where the dashed
(dotted) lines are at the jet surface (center), respectively, and the
heavy lines indicate the innermost contour of the high pressure region
(surface elliptical mode) or highest pressure region (first elliptical
body mode). The pressure and apparent emissivity plots indicate that
two bright filaments will accompany an elliptical distortion.  These
filaments will wind around the jet as the elliptical distortion rotates
spatially down the jet and can show brightness asymmetry as a result of
variation in the Doppler boosting factor induced by a varying flow
direction provided the observer is located at about the beaming angle
$\theta \approx 1/\gamma$.  At larger viewing angles this apparent
emissivity asymmetry is reduced.

Filaments produced by the first elliptical body mode will reside within
the jet and will dominate filaments produced by the surface elliptical
wave mode whose apparent emissivity falls rapidly inside the jet
surface.  The helical first body mode (not shown) would generate a
single bright twisted filament within the jet which would exhibit
apparent emissivity asymmetry, similar to that illustrated in Figure 3
for the elliptical mode filaments, as the filament winds around the
jet.  The helical surface mode which twists the entire jet is found to
result in little apparent emissivity variation \cite{h99b}. 

\section{Conclusion}

Kelvin-Helmholtz instability induced normal mode pinch structures have
been observed in axisymmetric non-relativistic and relativistic jet
simulations, and normal mode helical, elliptical, triangular and
rectangular structures have been observed in fully 3D non-relativistic
jet simulations.  In numerical simulations large amplitude helical and
elliptical normal mode structures have been found to lead to
significant mass entrainment. This mass entrainment slows ``light''
jets and prevents them from propagating to large distances while
remaining highly collimated.  The growth of large amplitude normal mode
structures can be slowed by high magnetosonic Mach numbers.  However,
magnetic acceleration and collimation schemes produce jets which pass
through a trans-Alfv\'enic and trans-magnetosonic region in which the
growth rate of the normal modes can be very large.  Numerical
simulations show that ``top hat'' profile ``light'' non-relativistic
jets will not survive this region. The helically twisted structures
that we observe in numerical simulations are on much shorter
spatial length scales than those of twisted structures seen at parsec
and kiloparsec length scales on extragalactic jets.

If magnetic jet acceleration and collimation schemes are to prove
viable for the production of observed jets that propagate to distances
orders of magnitude larger than the location of the Alfv\'en point,
they must flow through the transmagnetosonic region with sufficient
stability.  Additional stability both linearly and non-linearly may be
achieved by different density, temperature, magnetic and velocity
profiles, through rapid jet expansion \cite{h87} or by the embedding of
a jet in a surrounding fast wind.  In particular, a higher jet density
relative to the surrounding environment will slow the development of
instability.  The high jet density of non-relativistic protostellar
jets relative to the ambient and cocoon medium behind the jet bow shock
should allow the jets to propagate with sufficient stability through a
trans-Alfv\'enic region.  For galactic relativistic jets or
extragalactic highly relativistic jets magnetic acceleration schemes
would require that the Alfv\'en speed at the Alfv\'en point be near to
or greater than lightspeed in order to produce a relativistic jet.  In
this case, theory indicates that relativistic effects can be
stabilizing and lead to a scenario in which relativistic galactic and
extragalactic jets can propagate through the trans-Alfv\'enic region
with sufficient stability.

Once through the trans-Alfv\'enic region, continued jet expansion leads to
supermagnetosonic flow.  On the supermagnetosonic jet relativistic effects
serve to reduce the distortion amplitude of normal mode structures, and
this reduction in distortion amplitude should lead to a reduction in the mass
entrainment that slows and disrupts jet collimation.  Although
distortion amplitudes are reduced the predicted pressure fluctuations
combined with the Doppler boosting factor suggest that twisted
filamentary structures should be readily visible on relativistic jets.

\section*{Acknowledgements}

This work was supported by the National Science Foundation through
grants AST 9318397 and AST 9802955 to the University of Alabama and
through grant AST 9617032 to the University of Michigan. Numerical work
utilized the Pittsburgh Supercomputing Center and the Ohio Supercomputer
Center.

\end{document}